\documentclass[11pt]{article}
\usepackage{osid}
\usepackage{subcaption}
\usepackage{physics}
\usepackage{xcolor}
\usepackage{bbold}
\usepackage{graphicx}
\usepackage[T1]{fontenc}
\usepackage[utf8]{inputenc}
\usepackage{accents}
\usepackage{amsmath}
\usepackage{amssymb}
\usepackage{bm}

\newcommand{\ketom}{\ket{\Omega}}

\newcommand{\braom}{\bra{\Omega}}

\newcommand{\intom}{\int d \mu (\Omega)}

\newcommand{\phiphi}{\ket{\phi}\bra{\phi}}

\newcommand{\exval}[1]{\langle{#1}\rangle}

\newcommand\doublehat[1]{\bm{#1}}

\title{Nature and origin of the operators entering the master equation 
of an open quantum system}
\author{Giovanni Spaventa \\
{\footnotesize\it Institute of Theoretical Physics and IQST, Universit\"at Ulm, 
Albert-Einstein-Allee 11 D-89081, Ulm (Germany)
\& giovanni.spaventa@uni-ulm.de}\\[2ex]
Paola Verrucchi \\
{\footnotesize\it Istituto dei Sistemi
Complessi CNR, and Dipartimento di Fisica,
Universit\`a di Firenze, and INFN sezione di Firenze, via G.Sansone 1,
50019 Sesto Fiorentino (Italy)\\
\& verrucchi@fi.infn.it}}
\begin{document}

\maketitle

\begin{abstract}
By exploiting the peculiarities of a recently introduced formalism for 
describing open quantum systems (the Parametric Representation 
with Environmental Coherent States) we derive an equation of motion for 
the reduced density operator of an open quantum system that has the same
structure of the 
celebrated\\
Gorini–Kossakowski–Sudarshan–Lindblad equation, but holds regardless of markovianity being assumed.
The operators in our result have explicit expressions in terms of the 
Hamiltonian describing the interactions with the environment, and can be 
computed once a specific model is considered. We find that, instead of a 
single set of Lindblad operators, in the general (non-markovian) case 
there one set of Lindblad-like operators for each and every 
point of a symplectic manifold associated to the environment. This 
intricacy disappears under some assumptions (which are related to 
markovianity and the classical limit of the environment), under which it 
is possible to recover the usual master-equation formalism. Finally, we 
find such Lindblad-like operators for two different models of a qubit 
in a bosonic environment, and show that in the classical limit of the 
environment their renown master equations are recovered.
\end{abstract}

{\it\small
\begin{center}{\bf Prologue}\end{center}
\noindent I am not amongst the lucky persons who have been knowing 
Prof.~Kossakowski for decades. I knew his extraordinary works, 
of course, but I only met him in person 10 years ago, during the 44th 
Symposium on Mathematical Physics. 
It was my first time in Torun and I was nervous: 
I knew I would have presented my work to some of the most distinguished 
experts in OQS of the world, Prof.~Kossakowski in the first place.

\noindent
On the morning of the second day Andrzej gave his talk and 
my tension melt away like snow: he was authoritative and friendly, 
rigorous and talkative, focused and serene. He was a real master.
Suddenly I could not wait to tell him about my work. 

\noindent When I gave the talk, the day after, he was sitting in the 
first row of the lecture-hall and I simply felt proud and happy, because 
he was there. And this is how we got to know each other.
\vskip .2truecm
\noindent
Listening to Andrzej, looking him writing at the blackboard, following 
his extraordinary thoughts has been a pleasure. Talking with him a true 
honour.

\hfill Paola}

\section{Introduction}
\label{s.I}
Quantum information science, with its rapid growth over the past years, 
has been crucial in providing insights in the foundations of quantum 
mechanics, while allowing the development of quantum technologies. 
In fact, controlling a quantum device would be an 
impossible task without a deep understanding of the interplay between 
systems and their surrounding environments ~\cite{AvEtal20,SamachEtal21, 
ZhangEtal22,PassarelliFL22}. This is indeed one of the 
main goals in the analysis of Open Quantum Systems (OQS): to study the 
dynamics of quantum systems interacting with their equally quantum 
environments~\cite{Weiss99, BreuerP02, BenentiCS04, RivasH12}

One of the main features OQS, at variance with their isolated 
counterparts, is that they inherently exhibit memory effects, due to the 
dynamical generation of entanglement between the principal system and 
its environment. This means that their time evolution (in particular the 
time evolution of their density operator $\rho(t)$) is generally 
non-markovian, and therefore not described by differential equations. 
However, there are some cases in which entanglement between the two 
subsystems does not severely alter their dynamics, the best example 
being that of macroscopic environments, when their behaviour can be 
effectively described by classical dynamics~\cite{Schlosshauer07}. In 
such situations, time 
evolution in terms of a differential equation for $\rho(t)$ can be 
recovered, in the form of a so-called master equation, 
$\dot{\rho}=\mathcal{L}[\rho(t)]$, where $\mathcal{L}$ is dubbed 
generator of the master equation itself~\cite{BreuerP02,RivasH12}.

Thank to the work of Gorini, 
Kossakowski, Sudarshan \cite{GoriniKS76} and Lindblad \cite{Lindblad76}, 
it is known that $\mathcal{L}$ must be of a certain form if it has to 
describe markovian dynamics, thus defining the so-called GKSL 
equation~\cite{ChruscinskiP17}.
The operators entering such general expression (the Lindblad 
operators or lindbladians) are not derived from the microscopic details 
of the theory, and this makes it difficult to guess their form, and 
gives the GKSL result a phenomenological character it should not actually 
have. Moreover, the relevance of OQS dynamics in controlling quantum 
devices, such as the processors of quantum computers, makes it necessary 
to explicitly determine the lindbladians for the most diverse 
systems and environments
~\cite{Nakajima58,Zwanzig60,CaldeiraL83,BoulantEtal03,deLeeuwPP21,McDonaldC22}.

In this work we want to derive a GKSL-like master equation in a way such 
that the Lindblad operators emerge in terms of the actual interaction 
entering the model  under analysis. 
To this aim we use the formalism of Generalized Coherent States (GCS)
~\cite{ZhangFG90,Perelomov72,Perelomov86}, which is one of the best 
tools for describing the quantum-to-classical crossover, 
i.e. the way a quantum system may feature a classical behaviour when 
becoming macroscopic~\cite{Yaffe82,CCFVsoft20}.
In fact, referring to a recently introduced~\cite{CCGVpnas13} method to 
study OQS, namely the Parametric Representation with Environmental 
Coherent States (PRECS)~\cite{CCGVpnas13}, we exploit the peculiarities 
of GCS and 
describe the open system under analysis in terms of an 
ensemble of normalized pure states, parametrically dependent on the 
environmental configurations. This ensemble defines a
representation of $\rho(t)$ for which we obtain
a master equation  that
can be cast in the GKSL-like form, 
once an ansatz on the nature of some specific mathematical objects 
entering its derivation is made.
In this equation we recognize the operators playing the role of 
the lindbladians, and get their explicit form
in terms of the original hamiltonian $H$ of the total system,  so that 
$\dot{\rho}(t)$ can 
be written once $H$ is known or, alternatively, information 
on $H$ is obtained if $\rho(t)$ can be phenomenologically deduced, but 
$H$ cannot.

The structure of the paper is as follows: in Sec.~\ref{s.II}
 we outline the main features of the PRECS, which will serve as a basis 
for Sec.~\ref{s.III}, where we perform the 
time-derivative of the parametric representation of $\rho(t)$. 
Exploiting 
the known results about the dynamics of GCS, in Sec.~\ref{ss.IIIA} we 
show that the resulting equation is GKSL-like, provided that the 
nature of some mathematical objects appearing in its derivation is 
properly interpreted. The classical limit for the environment is 
considered in Sec.~\ref{ss.remarks}, while in Sec.~\ref{s.IV} we 
consider two models of qubits in bosonic environments, 
namely the {\it pure-dephasing} and the {\it Jaynes-Cummings} model,
in order to see how our results compare with what is already known.
Finally, in Sec.~\ref{s.V} we discuss about the 
additional information one can possibly obtain by our approach, 
and draw some conclusions.

\section{Parametric Representation with environmental coherent states} 
\label{s.II} 

The description of OQS in terms of reduced density 
operators, stems from an exact procedure that preserves the quantum 
character of the environment, namely that of tracing out the 
environmental degrees of freedom from the density operator of the total 
system. 

In fact, there exists another approach for dealing with OQS, based on 
the idea of considering them as if they were closed, i.e. described by 
an effective hamiltonian, whose time-dependent parameters (such as 
oscillating fields or fluctuating couplings) account for environmental 
effects. This approach implicitly assumes that the environment 
behaves classically, since the environmental operators are replaced by
time-dependent functions. This fact has two 
consequences: firstly, entanglement between the principal system and its 
environment has no place in this description (there cannot be 
entanglement between a quantum system and a classical one); secondly, by 
choosing ad-hoc time dependencies for the parameters in the effective 
hamiltonian, one completely neglects the environmental dynamics due to 
the interaction with the principal system, often dubbed 
\textit{back-action} \cite{CubittEW09,FCVpra16}.

In what follows we will rather adopt the recently introduced 
\cite{CCGVpnas13,CCGVosid13} Parametric 
Representation with Environmental Coherent States (PRECS), 
which provides a formally exact way to study the OQS dynamics in terms of a 
collection of pure states, each labelled by a set of parameters which 
are in one-to-one correspondence with (coherent) quantum states of the 
environment. In fact, the use of Generalized Coherent States (GCS) 
\cite{Perelomov72,Perelomov86,ZhangFG90} for describing the 
environment, naturally emerges in the study of the quantum-to-classical 
crossover~\cite{Yaffe82,Schlosshauer07}, that is, the limit in which a 
quantum system displays an 
effectively classical behaviour due to its being macroscopic. This 
implies that the PRECS provides an ideal formalism for studying the 
\textit{twilight zone} where an environment retains some of its quantum 
features, and yet starts being quite properly described by a 
classical-like theory~\cite{CCFVsoft20,Foti19}.

Let us briefly summarize the PRECS formalism, by considering a bipartite 
quantum system $\Psi = \Gamma \cup \Xi$, where $\Gamma$ and $\Xi$ play 
the roles of \textit{principal system} and \textit{environment}, 
respectively. The GCS are constructed for the environment $\Xi$, 
following the group-theoretical procedure of Ref.~\cite{ZhangFG90}, 
which 
we 
briefly outline below.

Starting from the hamiltonian of the environment,
$H_\Xi$,  one identifies the so called dynamical group $G$, i.e, the 
group of 
propagators that determine the evolution of $\Xi$. The procedure futher 
requires the 
choice of an arbitrary normalized element $\ket{R}$, called \textit{reference 
state}, in the 
Hilbert space $\mathcal{H}_\Xi$ of the environment, from which 
the identification of the \textit{maximum 
stability subgroup} of $G$ follows: this is the subgroup $F$ whose 
elements leave $\ket{R}$ unchanged, up to an irrelevant phase factor. 
From $F$, the coset $G/F$ and its associated symplectic manifold
${\cal M}$, whose points are in one-to-one correspondence with the 
elements of $G/F$~\cite{Lee12}, are defined, and each element of 
$G/F$ defines a state $\ketom \in \mathcal{H}_\Xi$. 
By construction, hence, GCS
are in one-to-one correspondence with points on ${\cal M}$. 
GCS for the environment will be hereafter dubbed {\it environmental 
coherent states} (ECS). These states are normalized but 
non-orthogonal, and they form an overcomplete 
set on 
$\mathcal{H}_\Xi$, i.e.
\begin{equation}
    \int_\mathcal{M}d\mu(\Omega)\ket{\Omega}\bra{\Omega}=\mathbb{1}_\Xi \, ,
    \label{eq: continuous_resolution}
\end{equation}
where $d\mu(\Omega)$ is a measure on $\mathcal{M}$ that is invariant 
under 
the action of elements in $G/F$.\\

Introducing the bases $\{\ket{k}\}$ and $\{\ket{\xi}\}$ of the Hilbert 
spaces $\mathcal{H}_\Gamma$ and $\mathcal{H}_\Xi$ respectively, any 
state of $\Psi$
\begin{equation}
    \ket{\psi} = \sum_{k\xi}c_{k\xi}\ket{k}\otimes\ket{\xi}\, ,
\end{equation}
can be expressed in terms of ECS thanks to the resolution of the 
identity \eqref{eq: continuous_resolution}:
\begin{equation}
\ket{\psi} = \sum_{k\xi}c_{k\xi} \ket{k} \otimes \int_\mathcal{M}d\mu(\Omega)\ket{\Omega}\bra{\Omega}\ket{\xi}
= \int_\mathcal{M}d\mu(\Omega) \chi(\Omega) \ket{\phi(\Omega)}\otimes\ket{\Omega} \, ,
\end{equation}
where
\begin{equation}
\begin{split}
\ket{\phi(\Omega)} = \frac{1}{\chi(\Omega)}\sum_k a_k(\Omega)\ket{k}~
~&,~~a_k(\Omega) = \sum_\xi c_{k\xi}\bra{\Omega}\ket{\xi}\,, \\
    \chi(\Omega)=&\sqrt{\sum_k |a_k(\Omega)|^2}\, .
\end{split}
\label{eq:coeff_def}
\end{equation}
The positive function $\chi^2(\Omega)$ is normalized on $\mathcal{M}$, i.e.
\begin{equation}
    \int_\mathcal{M} d\mu(\Omega) \chi^2(\Omega) = 1~,
\end{equation}
and can be interpreted as a probability distribution for $\Xi$ to be in 
the
state $\ket{\Omega}$ when the global state is $\ket{\psi}$. 

The density operator for the principal system, $\rho_\Gamma$, 
can be cast~\cite{CCGVpnas13} into the form
\begin{equation}
    \rho_\Gamma = \int_\mathcal{M} d\mu(\Omega) |\chi(\Omega)|^2 \ket{\phi(\Omega)}\bra{\phi(\Omega)}\, ,
    \label{eq:rhogamma}
\end{equation}
which means that the PRECS allows us to describe an open system $\Gamma$ 
in terms of normalized pure states $\ket{\phi(\Omega)}$, parametrically 
dependent on the environmental parameters $\Omega$. Each $\Omega$, on 
the other hand, is in one-to-one correspondence with ECS $\ket{\Omega}$,
the probability of whose occurrence is given by 
$\chi^2(\Omega)$.

\section{A time derivative}
\label{s.III}
The above expression \eqref{eq:rhogamma} above provides the reduced density operator 
of an OQS 
in terms of a probability distribution $\chi^2(\Omega)$ on the 
manifold $\mathcal{M}$. It is clear that, under the Markovian 
approximation, such density operator must satisfy a GKSL equation
\begin{equation}
    \dot{\rho}_\Gamma = -i[H_{\rm eff}, \rho_\Gamma] + \sum_k \gamma_k \Big( L_k \rho_\Gamma L_k^\dagger - \frac{1}{2}\{L_k^\dagger L_k, \rho_\Gamma\} \Big)
\end{equation}
for some $H_{\rm eff},\gamma_k$ and set of lindbladians $L_k$.

In this section we explicitly perform the time 
derivative of Eq.\eqref{eq:rhogamma}, with the goal 
of casting the resulting expression into a form that might allow the 
identification of the Lindblad operators in the Markovian limit.

For the sake of clarity, fron now on 
we will adopt the following conventions: the 
$\Omega$-dependencies, as well as the time ones, will be understood 
whenever possible, and restored when leading to a better understanding; 
moreover, in this section, and this one only, operators acting on
the whole composite system, 
i.e. on ${\cal H}_\Gamma\otimes{\cal H}_\Xi$, will appear
in bold, operators acting on ${\cal H}_\Gamma$ will be denoted by a hat, 
$\widehat{\cdot}$, while operators acting on ${\cal H}_\Xi$ will have no 
distinct sign;
hats will not be used for density operators and their time-derivatives.

Starting from Eq.\eqref{eq:rhogamma}, we consider the time derivative of 
$\rho_\Gamma$:
\begin{equation}
    \dot{\rho}_\Gamma = \int d\mu \bigg[ \widehat K+\chi^2(\Omega_t) 
\frac{d}{dt}\Big(\ket{\phi}\bra{\phi}\Big)\bigg],
    \label{eq:rhodot_K+phiphi} 
\end{equation}
where we have defined
\begin{equation}
\begin{split}
    \widehat K:=\dot{\chi}\ket{\phi}\bra{\phi}\chi^*  
+\chi\ket{\phi}\bra{\phi}\dot{\chi}^*\, .
\end{split}
\label{eq:kterm}
\end{equation}
The time derivative of $\chi(\Omega_t)$ is
\begin{equation}
\begin{aligned}[b]
   \dot{\chi}=&\frac{1}{2\chi}\frac{d}{dt}\sum_k | \sum_\xi 
c_{k\xi}|\exval{\Omega_t|\xi}|^2  \\ 
 =&\frac{1}{2\chi}\sum_{k\xi\xi'}\Big[ 
\Big( \dot{c}_{k\xi}c_{k\xi'}^*+c_{k\xi}\dot{c}_{k\xi'}^* \Big) 
\exval{\Omega_t|\xi}\bra{\xi'}\ket{\Omega}\\
 +&c_{k\xi}c_{k\xi'}^* \Big( 
\bra{\xi'}\ket{\Omega_t}\frac{d}{dt}\exval{\Omega_t|\xi} 
+ \exval{\Omega_t|\xi}\frac{d}{dt}\bra{\xi'}\ket{\Omega_t} \Big) \Big]\,;
\end{aligned}
\label{eq:dotchiom}
\end{equation}
the quantities $\frac{d}{dt}\exval{\Omega_t|\xi}$ require some care in 
being 
dealt with: they contain the dynamics of the parameter $\Omega$ on the 
manifold that, if $\Xi$ were isolated, would be given by the 
Hamilton-like equations of motion (see for instance 
Ref.~\cite{ZhangFG90}), with the function 
$H_\Xi(\Omega):=\exval{\Omega|H_\Xi|\Omega}$ playing the role of the 
classical hamiltonian.
On the other hand, $\Xi$ is not isolated, and the hamiltonian 
$\doublehat{H}$ with which we will be dealing acts upon 
${\cal H}_\Gamma\otimes{\cal H}_\Xi$: therefore, 
$\exval{\Omega|\doublehat{H}|\Omega}$ is
an operator acting on $\mathcal{H}_\Gamma$, and so is
$\frac{d}{dt}\exval{\Omega_t|\xi}$. This statement is not crystal clear 
from 
a mathematical viepoint, and we will get back to it in the subsection 
below; 
however, we anticipate that whenever an expression for
the above derivatives is explicitly available, it is indeed in the form 
of an operator for $\Gamma$.

Having said that, we define
\begin{equation}
    \begin{aligned}[b]
        a_k = \sum_\xi c_{k\xi}\exval{\Omega_t\xi} &\,, \quad b_k = 
\sum_\xi \dot{c}_{k\xi}\exval{\Omega_t|\xi} \\
       \text{and }\quad \widehat F_k = &\sum_\xi 
c_{k\xi}\frac{d}{dt}\exval{\Omega_t|\xi},
    \end{aligned}
    \label{e.definitions}
\end{equation}
and  get
\begin{equation}
\begin{aligned}[b]
    &\dot{\chi}=\frac{1}{2\chi}\sum_k \Big[a_k^*b_k+a_k 
b_k^*+a_k^*\widehat F_k+a_k \widehat F_k^\dagger \Big] \\ & = 
\frac{1}{2\chi}\sum_k 
\Big[r_k+a_k^* \widehat F_k+a_k \widehat F_k^\dagger \Big]\,,
\end{aligned}
\end{equation}
with $r_k\equiv 2\Re (a_k^*b_k)=a_k^*b_k+a_k b_k^*\,$.\\

\noindent Introducing the operators 
\begin{equation}
\widehat L_k = \frac{1}{a_k} (\widehat{\mathbb{1}} - a_k^* \widehat F_k )~,
\label{e.Lindbladians}
\end{equation}
it can be shown that $\widehat K$ in 
Eq.\eqref{eq:kterm} reads
\begin{equation}
\begin{aligned}[b]
    \widehat K=&\sum_k \big( r_k+1\big)\phiphi \\
-&\Big\{\frac{1}{2}\sum_k |a_k|^2
\left(\widehat L_k^\dagger \widehat L_k-\widehat F_k^\dagger \widehat F_k\right)
, \phiphi \Big\}\,,
\end{aligned}
\end{equation}
where $\{A,B\}=AB+BA$.\\
In order to get rid of the operators $\widehat F_k$, we use 
\begin{equation}
    \tr\rho_\Gamma = 1 \implies \tr\dot{\rho}_\Gamma = 0 \, ,
\end{equation}
implying
\begin{equation}
    \tr\int d\mu \bigg[  \widehat K + \chi^2\frac{d}{dt}\phiphi \bigg]=0\, 
.
\end{equation}

\noindent By the definition of $\widehat K$, and by making use of the cyclicity of the trace, the above condition can be rewritten as
\begin{equation}
\begin{aligned}[b]
    & \tr \int d\mu \bigg[\Big\{\frac{1}{2}\sum_k |a_k|^2 
\widehat F_k^\dagger 
\widehat F_k, \phiphi \Big\} \bigg] \\ & = \tr \int d\mu 
\bigg[-|\chi|^2\frac{d}{dt}\phiphi +\sum_k |a_k|^2 \widehat L_k \phiphi \widehat L_k^\dagger - \sum_k (r_k+1)\phiphi \bigg]\,.
\end{aligned}
\end{equation}

What we have obtained above is an equation of the form $\tr A = \tr B$, which tells us that the operators $A$ and $B$ are equal up to a traceless operator $C$ (which is, in fact, simply their difference $C=A-B$). \\
Now, using this fact into the expression for $\dot{\rho}_\Gamma$ we get
\begin{equation}
    \begin{aligned}[b]
        \dot{\rho}_\Gamma = \widehat C + \int d\mu \bigg[ \sum_k |a_k|^2 
\bigg(  
\widehat L_k \ket{\phi}\bra{\phi} \widehat L_k^\dagger - \frac{1}{2} \Big\{ \widehat L_k^\dagger \widehat L_k, \ket{\phi}\bra{\phi} \Big\} \bigg) \bigg]\,.
    \end{aligned}
    \label{eq:intermediate_result}
\end{equation}

\noindent Let us now introduce the operator $\widehat R(\Omega)$, defined as
\begin{equation}
    \widehat R(\Omega) = 
\chi^2(\Omega)\ket{\phi(\Omega)}\bra{\phi(\Omega)} \, ,
\label{e.R}
\end{equation}
and such that $\rho_\Gamma = \intom \widehat R(\Omega)$. 
In terms of $\widehat R(\Omega)$, we can write
\begin{equation}
    \begin{aligned}[b]
        \dot{\rho}_\Gamma = \widehat C + 
\int d\mu \bigg[ \sum_k \gamma_k \bigg(  \widehat L_k \widehat R \widehat L_k^\dagger - 
\frac{1}{2} \Big\{ \widehat L_k^\dagger \widehat L_k, \widehat R \Big\} \bigg) \bigg]\,,
    \end{aligned}
\end{equation}
in whose second term we recognize the structure of 
the dissipator as in the GKSL equation, 
with $\gamma_k = \abs{\frac{a_k}{\chi}}^2=\abs{\bra{k}\ket{\phi(\Omega)}}^2$.\\

By comparing the results above with the GKSL equation, it is clear that 
the operator $\widehat C$ must be related to the unitary term 
$-i[\widehat H_{\rm eff}, \rho_\Gamma]$, 
and this can be further justified by the fact that 
$\tr \widehat C=0$ and every traceless operator can be written as a 
commutator, hence we take $\widehat H_{\rm eff}$ such that
\begin{equation}
    \widehat C = -i \Big[ \widehat H_{\rm eff}, \rho_\Gamma \Big] \, .
\end{equation}

\noindent Finally, the equation for $\dot{\rho}_\Gamma$ reads
\begin{equation}
        \dot{\rho}_\Gamma =-i\big[\widehat H_{\rm eff},\rho_\Gamma \big] + 
\int d\mu\sum_k \gamma_k \bigg(  \widehat L_k \widehat R \widehat L_k^\dagger - 
\frac{1}{2} \Big\{ \widehat L_k^\dagger \widehat L_k, \widehat R \Big\} \bigg)\, ,
        \label{eq:result_integral}
\end{equation}
which is indeed of the GKSL form with the Lindblad operators 
as from Eqs.\eqref{e.Lindbladians}, with their explicit form and actual 
nature as operators acting on ${\cal H}_\Gamma$ considered 
in the following subsection.

\subsection{Time evolution of ECS}
\label{ss.IIIA}
Let us get back to one of the key definitions in the above derivation, 
namely that of $\frac{d}{dt}\bra{\xi}\ket{\Omega_t}$ introduced in 
Eq.\eqref{eq:dotchiom}. We have consistently considered 
these objects as operators, which is ultimately the reason why the 
$\Omega$-dependent lindbladians $\widehat L_k(\Omega)$ are operators on 
$\mathcal{H}_\Gamma$. We 
now clarify this claim, resorting to the well-known dynamical 
properties of GCS, as found, for instance, in Ref.~\cite{ZhangFG90}.

The theory of GCS defines a bijective 
map between states $\ketom$ and points $\Omega$ on 
$\mathcal{M}$, whose coordinates dynamically evolve on ${\cal M}$ 
according to the Hamilton equations of motion 
\begin{equation}
     \dot{\Omega} = -i\frac{\partial H(\Omega)}{\partial \Omega^*} \quad,\qquad \dot{\Omega}^* = +i\frac{\partial H(\Omega)}{\partial \Omega}\,,
     \label{eq:hamilton}
\end{equation}
with $H(\Omega):=\braom H \ketom$ if $\Xi$ were isolated; therefore, it seems 
reasonable to associate a time evolution to the ECS,
exploiting the fact that there is one defined on the manifold, 
$\ket{\Omega_t}=\ket{\Omega(t)}$.

In particular, for an isolated system with GCS $\ket{\Omega}$, the 
inner product $\bra{\xi}\ket{\Omega}$ is a function $\phi_\xi(\Omega)$ 
on ${\cal M}$ with time derivatives
\begin{equation}
\begin{aligned}[b]
    & \frac{d\phi_\xi}{dt} = \frac{\partial \phi_\xi}{\partial \Omega}\dot{\Omega} + \frac{\partial \phi_\xi}{\partial \Omega^*}\dot{\Omega}^*\\ 
    & = -i \frac{\partial \phi_\xi}{\partial \Omega}\frac{\partial H(\Omega)}{\partial \Omega^*} +i \frac{\partial \phi_\xi}{\partial \Omega^*} \frac{\partial H(\Omega)}{\partial \Omega} = -i \{ \phi_\xi(\Omega), H(\Omega) \} \,,
\end{aligned}
\end{equation}
where
\begin{equation}
    \{f,g\}_{\rm P} 
= \frac{\partial f}{\partial \Omega}\frac{\partial g}{\partial \Omega^*} - \frac{\partial f}{\partial \Omega^*}\frac{\partial g}{\partial \Omega}
\end{equation}
is the Poisson bracket on $\mathcal{M}$. Therefore, for an isolated 
system, the time evolution of $\phi_\xi(\Omega)$ is dictated by the 
Poisson bracket of $\phi_\xi(\Omega)$ and the function $H(\Omega)=\bra{\Omega}H\ket{\Omega}$.\\

The problem with which we are dealing, though, is that of a composite system $\Psi = \Gamma \cup \Xi$, with a generic hamiltonian
\begin{equation}
    \doublehat{H} = \sum_i g_i \widehat O_i^\Gamma \otimes O_i^\Xi 
\end{equation}
where $g_i$ are the couplings, while $\widehat O_i^\Gamma$ and $O_i^\Xi$ are 
operators on $\mathcal{H}_\Gamma$ and $\mathcal{H}_\Xi$, respectively.\\
This means that, once the ECS are constructed for $\Xi$, 
$\bra{\Omega} \doublehat{H}\ket{\Omega}:=\widehat H(\Omega)$ is not a function 
but rather an operator on $\mathcal{H}_\Gamma$, in fact:
\begin{equation}
    \widehat H(\Omega) = \sum_i g_i \bra{\Omega} O_i^\Xi \ket{\Omega} \widehat 
O_i^\Gamma \equiv\sum_i g_i \theta_i(\Omega) \widehat O_i^\Gamma \, ,
\end{equation}
where we have introduced the expectation values 
\begin{equation}
    \theta_i(\Omega) \equiv \bra{\Omega} O_i^\Xi \ket{\Omega}\,. 
\end{equation}

\noindent Following the above reasoning, though, we can still refer to 
Eq.\eqref{eq:hamilton} and write:
\begin{equation}
\begin{aligned}[b]
    & \widehat{\dot{\phi}}_\xi:={\frac{d\phi_\xi}{dt}} = -i \bigg( 
\frac{\partial 
\phi_\xi}{\partial \Omega}\frac{\partial \widehat H(\Omega)}{\partial \Omega^*}
- \frac{\partial \phi_\xi}{\partial \Omega^*} \frac{\partial 
\widehat H(\Omega)}{\partial \Omega}\bigg) \\ 
    & = -i \sum_i g_i \Big\{ \phi_\xi (\Omega) , \theta_i(\Omega) \Big\}_{\rm P} 
\widehat O_i^\Gamma  \,:
\end{aligned}
\end{equation}
\noindent getting back to Eq.~\eqref{e.definitions}, we can 
then explicitly write
\begin{equation}
\begin{aligned}[b]
    &\widehat  F_k(\Omega) = -i\sum_\xi \sum_i g_i c_{k\xi} 
\Big\{\bra{\Omega}\ket{\xi} , \theta_i(\Omega) \Big\}_{\rm P} \widehat 
O_i^\Gamma \\ 
    & = -i \sum_i g_i \Big\{ \sum_\xi c_{k\xi}\bra{\Omega}\ket{\xi} , 
\theta_i(\Omega) \Big\}_{\rm P} O_i^\Gamma  \\
& = -i \sum_i b_i^{(k)}(\Omega) \widehat O_i^\Gamma \, ,
\label{e.Fk}
\end{aligned}
\end{equation}
with
\begin{equation}
    b_i^{(k)}(\Omega) = g_i \big\{ a_k(\Omega) , \theta_i(\Omega) 
\big\}_{\rm P} \, .
    \label{eq:coeff_bik}
\end{equation}
Thanks to the above results
one obtains, via Eq.~\eqref{e.Lindbladians},  the explicit form of the 
operators playing the role of the lindbladians in our GKSL-like 
equation \eqref{eq:result_integral}. 
We underline that, as mentioned in the Introduction, these operators
depend on $\Omega$, from whom they inherit an essential time dependence.

\subsection{Remarks on Markovianity and the Classical Limit}
\label{ss.remarks}

The integro-differential equation \eqref{eq:result_integral} 
evidently resembles a GKSL equation. though it
does not embody any markovian approximation. Consistently it cannot be 
considered a genuine master equation, as it turns clear after noticing, 
for instance, 
that the integral on its r.h.s. contains the operator 
$\widehat R=\widehat R(\Omega_t)$  rather than 
$\rho_\Gamma(t)$. However, reminding that 
$\widehat R$ is proportional to $\chi^2(\Omega)$ by definition \eqref{e.R}, 
one can guess that conditions upon the probability distribution of ECS 
on ${\cal M}$, embodied by $\chi^2(\Omega)$ itself, can transform 
Eq.\eqref{eq:result_integral} into some more familiar, purely 
differential, equation.
In fact, it can be demonstrated
~\cite{Yaffe82, CCGVpnas13, FotiEtal19, CCFVsoft20} 
that taking the classical limit for 
$\Xi$, and $\Xi$ only, implies
\begin{equation}
\chi^2(\Omega)\to\sum_i p_i \delta(\Omega-\Omega_i)~,
\label{e.chi2_classical}
\end{equation}
with $p_i$ positive coefficients such that $\sum_i p_i=1$,
and 
\begin{equation}
    \bra{\Omega_i}\ket{\Omega_j}\to\delta_{ij}\,\,\forall t\,\,~,~\,\, 
\bra{\xi}\ket{\Omega_i}\bra{\Omega_i}\ket{\xi'}\to\delta_{\xi\xi'}\,\,\forall 
t\forall i\, ,
    \label{eq:conds}
\end{equation}
where $\to$ indicates the classical limit of the environment.
If this is the case, it is
\begin{equation}
    \rho_\Gamma(t)\to \sum_i p_i \ket{\phi(\Omega_i)}\bra{\phi(\Omega_i)} \, ,
\end{equation}
with $\bra{\phi(\Omega_i(t)}\ket{\phi(\Omega_j(t))}\to \delta_{ij}$, and hence, defining 
\begin{equation}
    P_i = \ket{\phi(\Omega_i)}\bra{\phi(\Omega_i)}\,,
\end{equation}
one gets the following equation of motion for $\rho_\Gamma$:
\begin{equation}
    \dot{\rho}_\Gamma\to  -i\Big[ \widehat H_{\rm eff} , \rho_\Gamma \Big] + 
\sum_{ki}  p_i\gamma_{ki} \bigg( \widehat F_{ki}\widehat P_i \widehat F_{ki}^\dagger 
- \frac{1}{2}\Big\{\widehat F^\dagger_{ki}\widehat F_{ki}, \widehat P_i \Big\} 
\bigg) \, ,
    \label{eq:decoupled_markov}
\end{equation}
where $\widehat F_{ki}\equiv \widehat F_k(\Omega_i(t))$ and 
$\gamma_{ki}\equiv\gamma_k(\Omega_i(t))$.\\
Under the condition 
\begin{equation}
    \sqrt{\gamma_{ki}}F_{ki}\equiv\sqrt{\gamma_{k}}F_{k} \quad \forall i\,,
\end{equation}
the equation becomes
\begin{equation}
    \dot{\rho}_\Gamma\to -i\Big[ \widehat H_{\rm eff} , \rho_\Gamma \Big] + 
\sum_{k}  \gamma_{k} \bigg( \widehat F_{k}\rho_\Gamma \widehat F_{k}^\dagger - 
\frac{1}{2}\Big\{\widehat F^\dagger_{k}\widehat F_{k}, \rho_\Gamma \Big\} \bigg) 
\, ,
\label{e.GKSLclassical}
\end{equation}
which is a genuine GKSL equation for $\rho_\Gamma$; moreover, if the 
parameters $\Omega_i$ do not vary in time, so do the operators 
$\widehat F_k$, that are hence recognized as true lindbladians.
This result confirms that there are conditions leading to a markovian 
master equation for $\rho_\Gamma(t)$: they are are embodied in Eqs.~
\eqref{e.chi2_classical} and \eqref{eq:conds}, which are 
demonstrated to hold \cite{Foti15}, when $\Xi$ becomes macroscopic and 
can be effectively described in the classical formalism, i.e. for a 
classical environment.

Before moving to the next section, we notice that whether we consider 
the operators $\widehat F_k$ or their siblings $\widehat L_k$ as 
lindbladians does 
not make much difference, as Eq.~\ref{e.GKSLclassical} can be written using 
these or the others by properly redefining $\widetilde H_{\rm eff}$ 
\cite{Spaventa19}.

\section{A qubit in a bosonic environment}
\label{s.IV}
Having found a GKSL-like equation for the time derivative of the density 
operator of an OQS in the parametric representation, we now want to see 
how this result works for two specific models and, in particular, 
derive an explicit form for the operators $\widehat F_k$ playing the 
role of lindbladians.
Both models describe a qubit $\Gamma$ interacting with a bosonic 
environment $\Xi$: the 
first one is the so-called {\it pure-dephasing} model, while the second 
is 
the well-known Jaynes-Cummings model.

For the sake of a lighter notation, and given that the formalism should
now be clear, in this section we drop any distinctive signs for
operators, no matter upon which Hilbert space they act.

\subsection{Pure-Dephasing Model}

By \textit{pure-dephasing model} it is usually meant one in which
the operators acting on $\Gamma$ that enter the total hamiltonian, 
commute 
with each other: this feature reflects into the existence of a preferred 
basis in $\mathcal{H}_\Gamma$, hereafter indicated by 
$\{\ket{\gamma}\}_{\mathcal{H}_\Gamma}$, which simultaneously 
diagonalizes all of the above operators. The model's dynamics is often 
referred to as \textit{off-diagonal}, since the diagonal elements of 
$\rho_\Gamma(t)$ with respect to the preferred basis are constant in 
time.

Introducing the Pauli matrices $\sigma_x,\sigma_y,\sigma_z$ to describe 
the qubit, and the 
creation,annihilation operators $a^\dagger,a$ for the bosonic 
environment,  
the total hamiltonian, with $\hbar=1$, is
\begin{eqnarray}
    H &=& \mathbb{1} \otimes \omega a^\dagger a + g \sigma_z \otimes 
\big( a + a^\dagger \big)
    \label{eq:pdham_1}\\
&=& H_\Xi+H_{int}~,
\end{eqnarray}
where
\begin{equation}
\quad H_\Xi=\omega a^\dagger a\,, \quad 
H_{int}=g\sigma_z \otimes (a+a^\dagger)\,.
\label{e.splitH}
\end{equation}
In this case the proper ECS are the usual Glauber coherent states 
\begin{equation}
    \ket{\alpha}=D(\alpha)\ket{0},
\end{equation}
with $\alpha\in\mathbb{C}$ and 
\begin{equation}
    D(\alpha)=e^{\alpha a^\dagger - \alpha^* a}\,,
\end{equation}
often called {\it displacement operator}.
These coherent states are such that
\begin{equation}
\begin{aligned}[b]
    & a\ket{\alpha}=\alpha\ket{\alpha} \quad, \qquad \ket{\alpha}=e^{-|\alpha|^2/2}\sum_n \frac{\alpha^n}{\sqrt{n!}}\ket{n}, \\
    & \bra{\beta}\ket{\alpha}=e^{-\frac{1}{2}(|\alpha|^2+|\beta|^2)}e^{\beta^*\alpha},
\end{aligned}
\end{equation}
where $\ket{n}$ are the Fock states (i.e. eigenstates of $a^\dagger a$), 
while the manifold $\mathcal{M}$ is the complex plane, with the invariant measure
\begin{equation}
    d\mu(\alpha)=\frac{d\alpha d\alpha^*}{\pi} \,.
\end{equation}

We now move to the derivation of an explicit form for the operators 
$F_k(\alpha)$ for this model:
we have 
\begin{equation}
    \frac{\partial}{\partial \alpha^*}H(\alpha) = \frac{\partial}{\partial \alpha^*} \Big( \omega|\alpha|^2\mathbb{1} + g(\alpha+\alpha^*)\sigma_z \Big) = g\sigma_z+\omega \alpha\mathbb{1}\,,
\end{equation}
and
\begin{equation}
    D \left (\dot{\alpha} \right) = \exp \Big[-i(g\sigma_z+\omega\alpha\mathbb{1})\otimes a^\dagger -i(g\sigma_z+\omega\alpha^*\mathbb{1})\otimes a \Big]\,.
\end{equation}
To compute the explicit action of the operator above on the reference 
state, we write it as 
\begin{equation}
    D(\dot{\alpha}) = e^{-i(Y+\bar{Y})}
\end{equation}
with
\begin{equation}
    Y=g\sigma_z+\omega\alpha\mathbb{1}\otimes a^\dagger\, ,\quad \bar{Y}=g\sigma_z+\omega\alpha^*\mathbb{1}\otimes a,
\end{equation}
and notice that
\begin{equation}
\begin{aligned}[b]
    & [Y,\bar{Y}] = -\Big(g^2+\omega^2|\alpha|^2\Big)\mathbb{1}\otimes\mathbb{1}-g\omega \Big(\alpha+\alpha^*\Big)\sigma_z\otimes\mathbb{1}\,, \\
    & \big[ Y, \big[ Y , \bar{Y} \big] \big] = \big[ \bar{Y}, \big[ Y , \bar{Y} \big] \big] = 0\,,
\end{aligned}
\end{equation}
implying that the Baker-Campbell-Hausdorff 
formula provides
\begin{equation}
    e^{-i(Y+\bar{Y})} = e^{-iY}e^{-i\bar{Y}}e^{-\frac{1}{2}[Y,\bar{Y}]} \, .
\end{equation}
The expression for $F_\pm(\alpha)$ consequently reads
\begin{equation}
    F_\pm(\alpha) = d_\pm(\alpha)\mathbb{1}+b_\pm(\alpha)\sigma_z \, ,
    \label{eq:F_pd}
\end{equation}
where $d_\pm(\alpha)$ and $b_\pm(\alpha)$ are functions of $\alpha$ 
whose expression is rather involved~\cite{FCVpra16,Spaventa19}, but gets 
simplified when the 
classical limit for the environment is taken. In particular, in such 
limit we find 
(assuming the initial state is
$\frac{1}{\sqrt{2}}(\ket{+}+\ket{-})\otimes\ket{\Xi}$ for some state 
$\ket{\Xi}$ of the environment)
\begin{equation}
\begin{aligned}[b]
    & d^{C}_\pm(\alpha) = \frac{1}{\sqrt{2}}e^{-\frac{1}{2}|\beta|^2}e^{\mp i\omega\beta\alpha^*}\cos{g\beta}\, , \\
    & b^{C}_\pm(\alpha) = \mp  \frac{i}{\sqrt{2}}e^{-\frac{1}{2}|\beta|^2}e^{\mp i\omega\beta\alpha^*}\sin{g\beta}\, ,
\end{aligned}
\end{equation}
where
\begin{equation}
    \beta(t) =  \frac{g}{\omega}(1-e^{-i\omega t}) \, .
\end{equation}
Consequently, Eq.~\eqref{eq:result_integral} takes the form
\begin{equation}
    \dot{\rho}_\Gamma =   -i [H_{\rm eff} , \rho_\Gamma] + \int d\mu\,\Gamma \,\Big( \sigma_z R \sigma_z - R \Big) 
\end{equation}
where
\begin{equation}
    H_{\rm eff} = \int d\mu \frac{1}{\chi^2} \sum_{k=\pm}\Im\Big(a_kb_k+|a_k|^2d^*_kb_k \Big) \sigma_z = h\sigma_z \,
\end{equation}
is the free hamiltonian of a qubit in an external time-dependent magnetic field pointing in the $z$ direction, while
\begin{equation}
    {\rm T} := \sum_{k=\pm} \gamma_k |b_k|^2 = 
\frac{1}{2}e^{-|\beta|^2}|\sin(g\beta)|^2 e^{-2\omega\Im (\alpha^*\beta)}.
\end{equation}
Since $R=\chi^2\phiphi$ contains the distribution $\chi^2$, which is 
significantly nonzero only in a small region around $\alpha=\pm\beta(t)$ (where $\Im(\beta^*\alpha)\approx 0$), we can safely take T out of the integral, use $\int d\mu R=\rho_\Gamma$, and write
\begin{equation}
    \dot{\rho}_\Gamma = -i\big[H_{\rm eff},\rho_\Gamma\big]+ \Gamma\Big( \sigma_z \rho_\Gamma \sigma_z - \rho_\Gamma \Big)\,,
\end{equation}
where
\begin{equation}
    {\rm T}=\frac{1}{2}e^{-|\beta|^2}|\sin(g\beta)|^2\,.
\end{equation}

\begin{figure}[h]
\begin{subfigure}{.5\textwidth}
  \centering
  \includegraphics[width=1\linewidth]{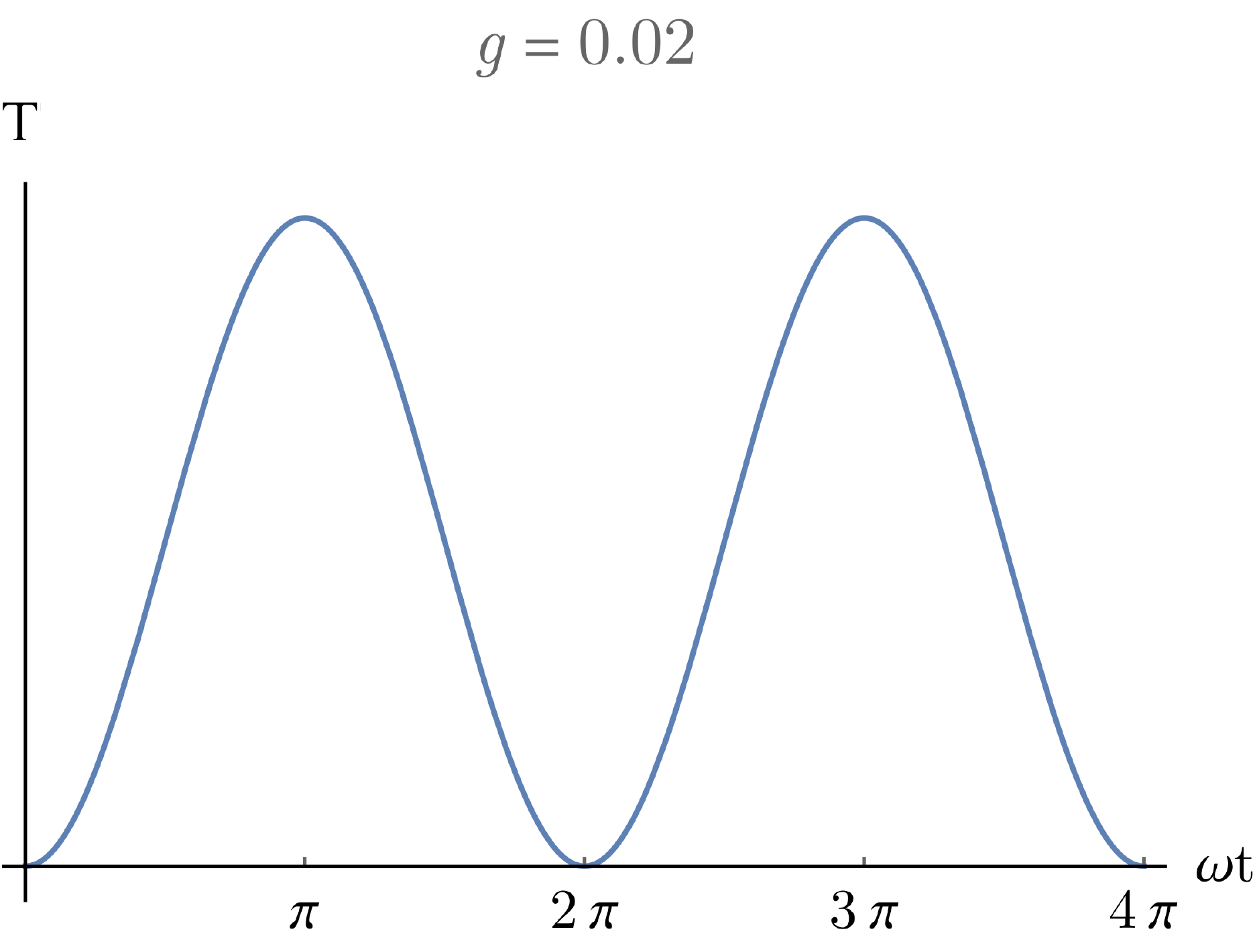}
\end{subfigure}
\begin{subfigure}{.5\textwidth}
  \centering
  \includegraphics[width=1\linewidth]{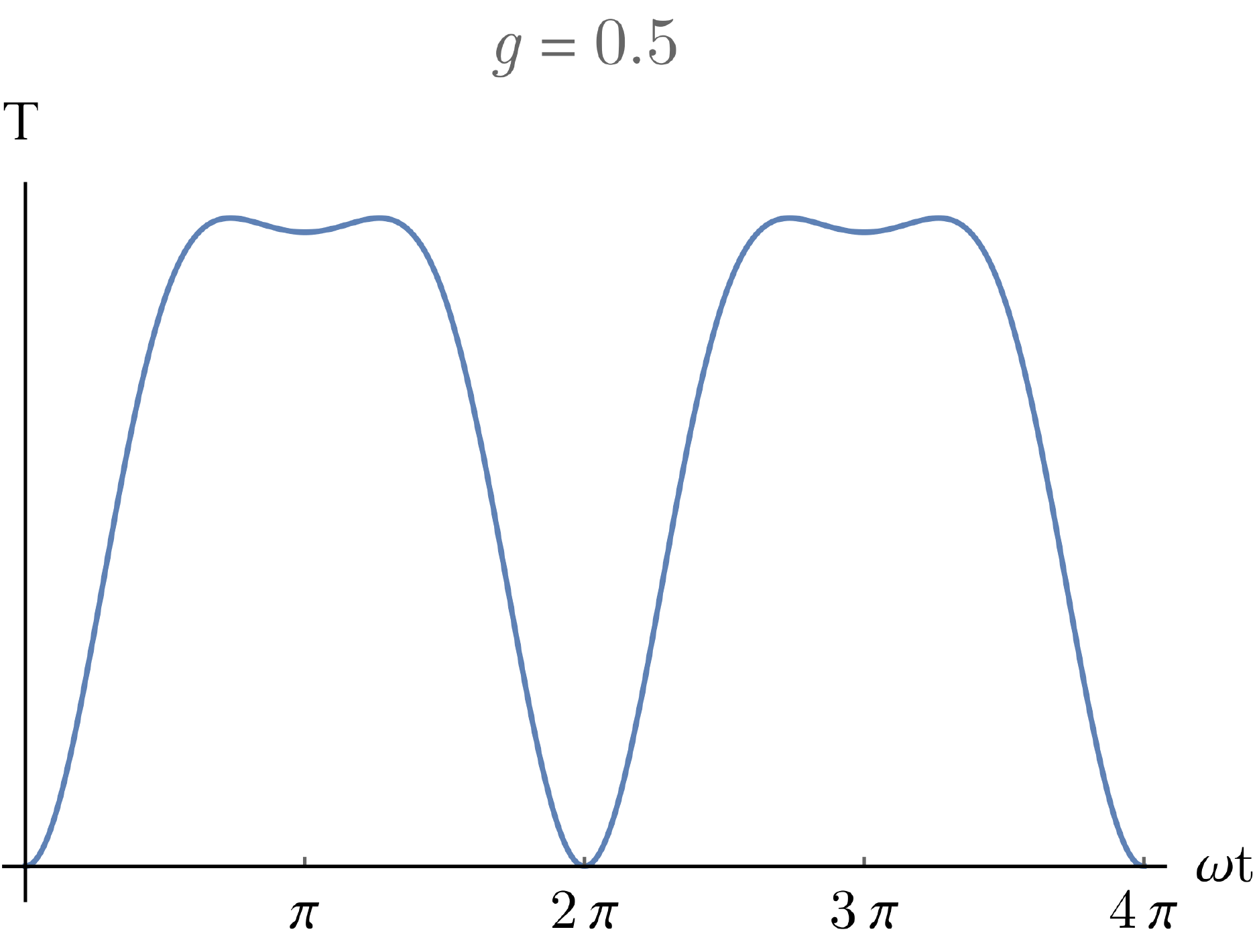}
\end{subfigure}

\begin{subfigure}{.5\textwidth}
  \centering
  \includegraphics[width=1\linewidth]{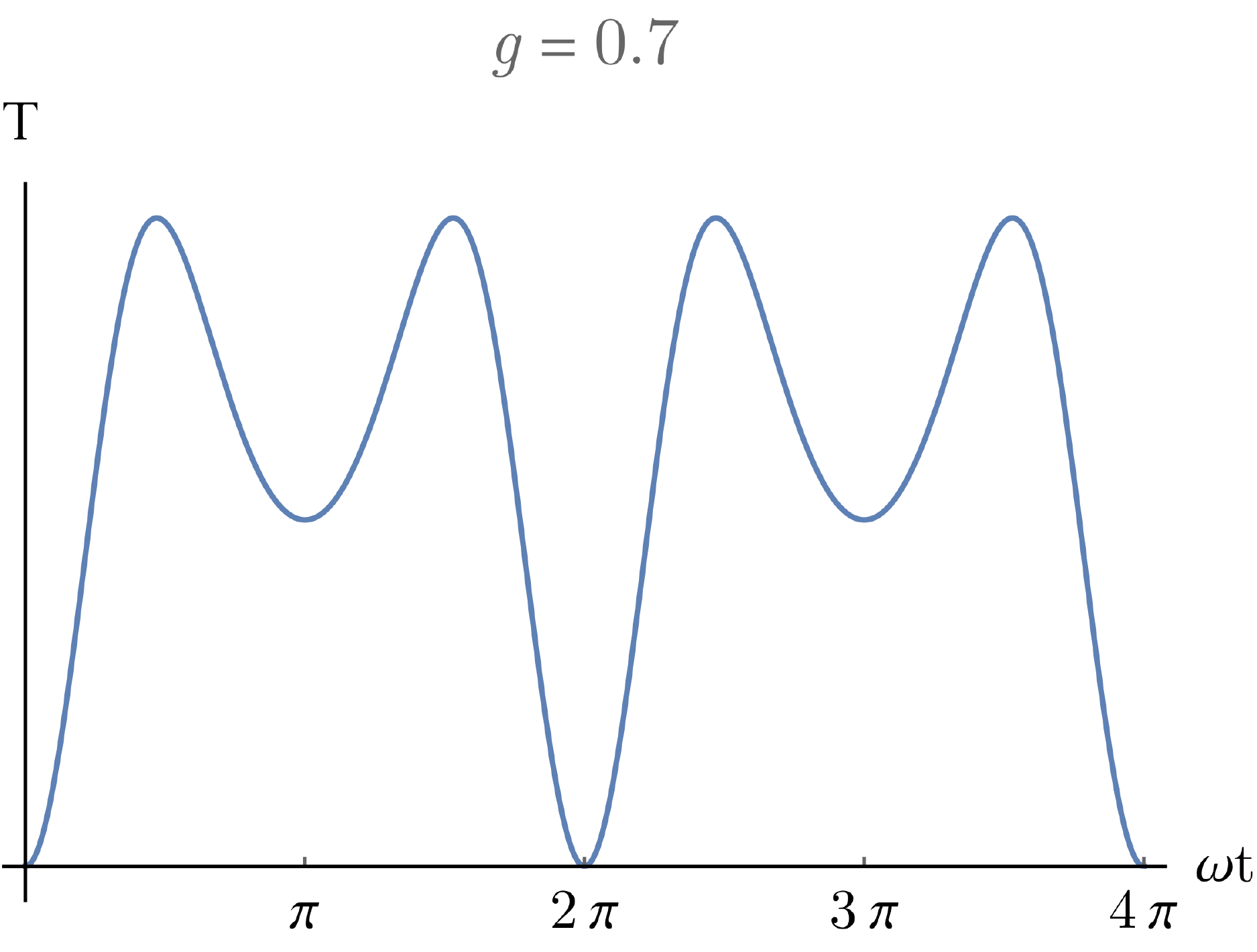}
\end{subfigure}
\begin{subfigure}{.5\textwidth}
  \centering
  \includegraphics[width=1\linewidth]{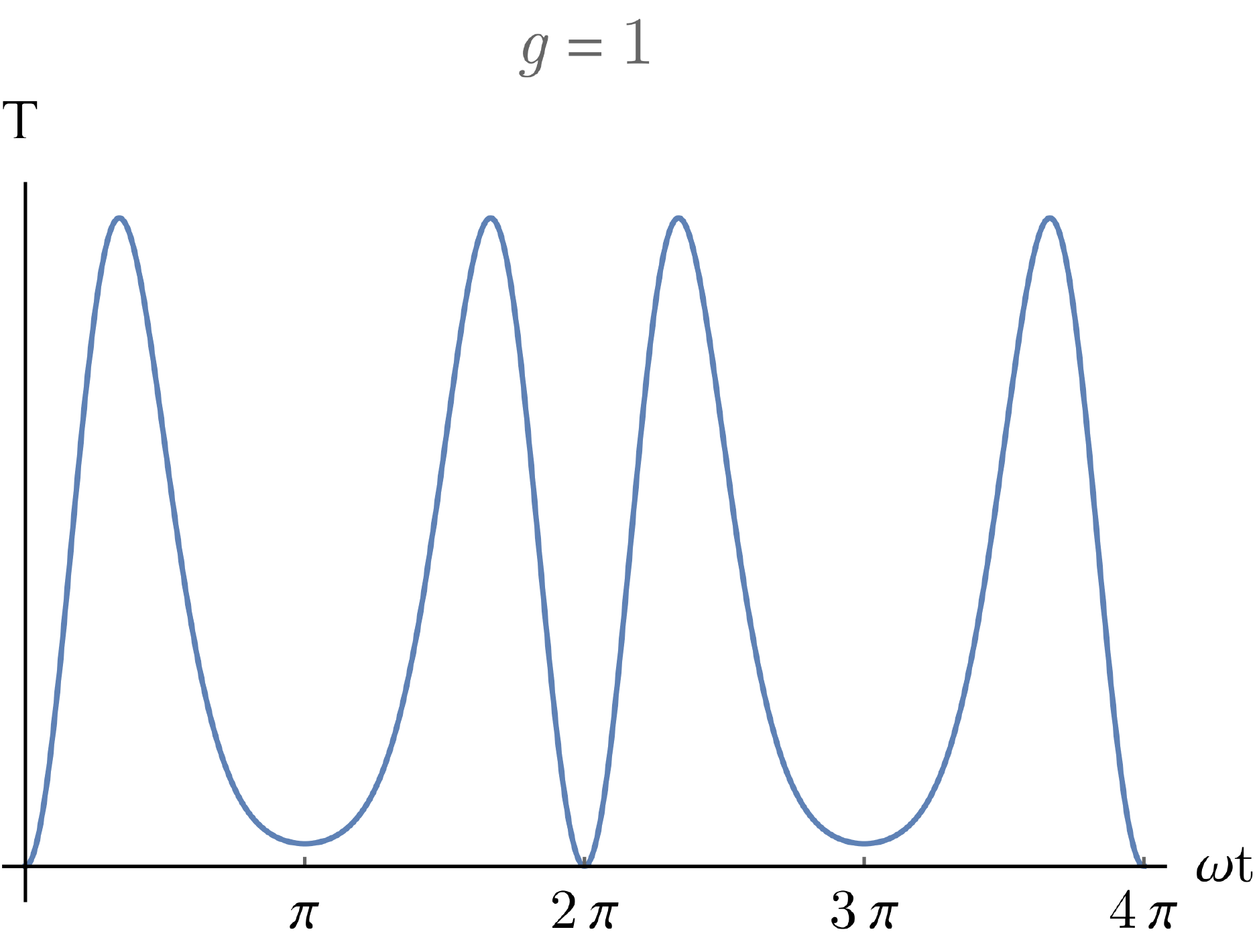}
\end{subfigure}\\
\begin{subfigure}{.5\textwidth}
  \centering
  \includegraphics[width=1\linewidth]{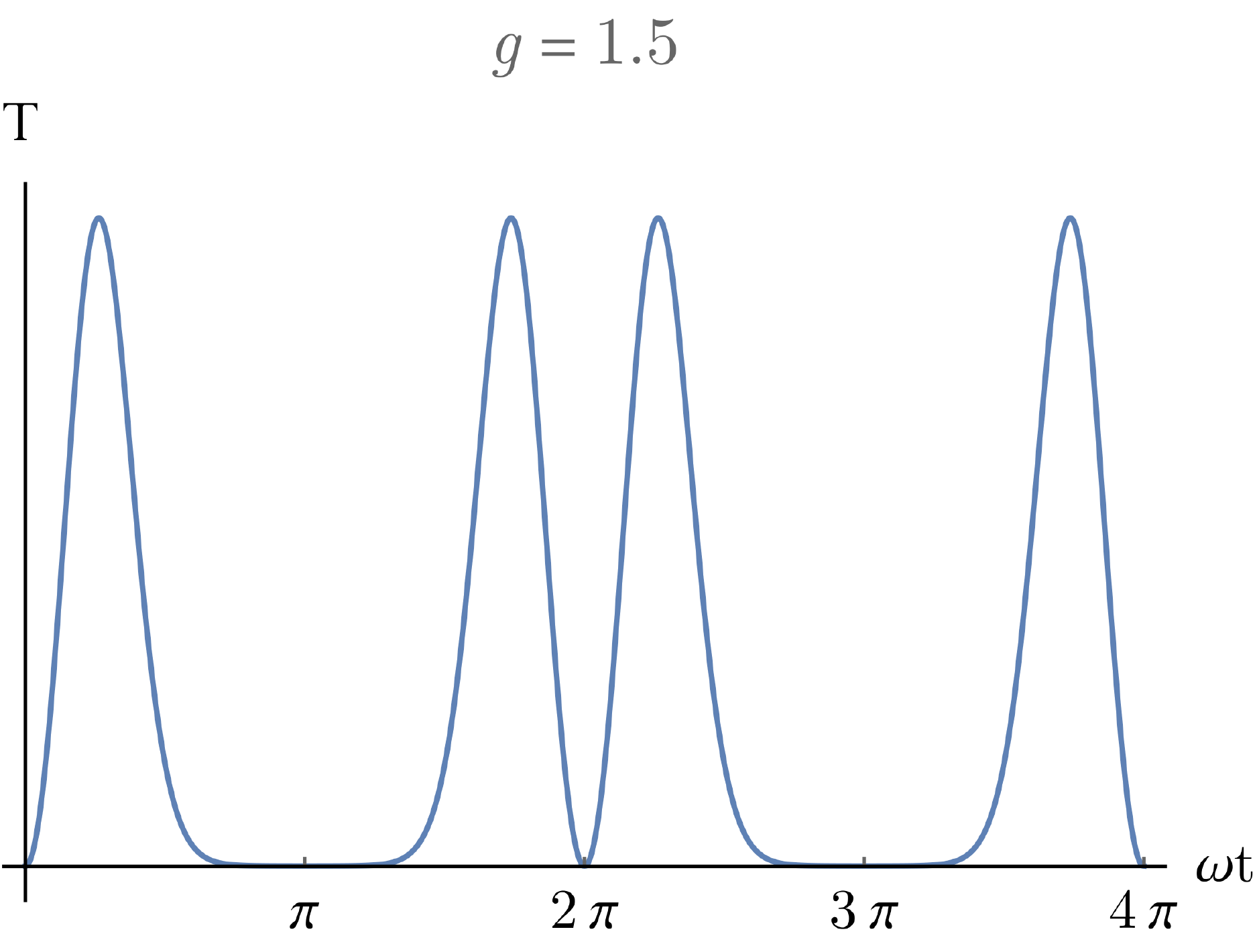}
\end{subfigure}
\begin{subfigure}{.5\textwidth}
  \centering
  \includegraphics[width=1\linewidth]{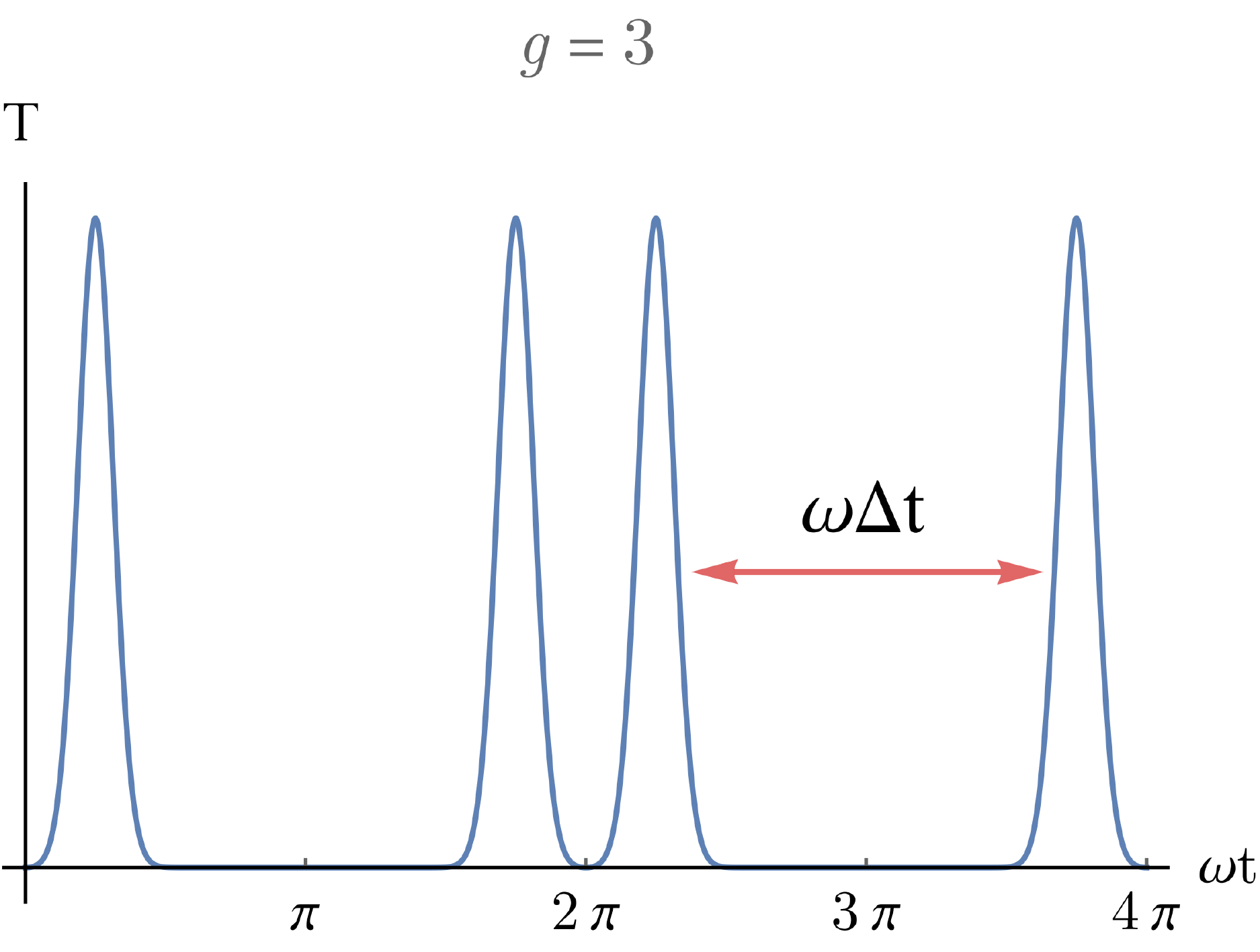}
\end{subfigure}
\caption{Plot of ${\rm T}(t)$ for several values of the coupling $g$. 
The functions are normalized to $1$.}
\label{fig:Gamma(t)}
\end{figure}
\noindent For small values of $g$ it is
\begin{equation}
    {\rm T}(t) \sim \frac{1}{2}g^2|\beta(t)|^2 = 
\frac{g^4}{\omega^2}(1-\cos(\omega t))\, ;
\end{equation}
however, as $g\to\infty$, the interval of time $\Delta t$ during which 
the function T$(t)$ is approximately zero, grows larger and larger 
in proportion to its period. This can be seen in 
Fig.\ref{fig:Gamma(t)} and it means that there are large intervals of 
time during which T$(t)\approx 0$ and the system evolves unitarily 
under the effective hamiltonian $H_{\rm eff}$, which, we recall, is the 
free hamiltonian of a qubit in an external time-dependent magnetic 
field.

In other words, in the strong coupling regime $g\to\infty$, there is no 
decoherence and the qubit evolves as if it were a closed system, under 
the influence of an effective hamiltonian containing a time-dependent 
field.\\

\subsection{Jaynes-Cummings Model}

Let us now consider another model for a qubit in a bosonic environment, 
namely the Jaynes-Cummings model. This model is defined by the 
hamiltonian
\begin{equation}
    H=\mathbb{1}\otimes \omega a^\dagger a + g(\sigma^+ \otimes a + 
\sigma^- \otimes a^\dagger )=H_\Xi+H_{int}~,
\label{e.JC}
\end{equation}
where
\begin{equation}
\quad H_\Xi=\omega a^\dagger a\,, \quad 
H_{int}=g\left(\sigma^+\otimes a+\sigma^-\otimes a^\dagger\right)~.
\label{e.splitHJC}
\end{equation}
The operators acting on $\Gamma$, $\sigma^+$ and $\sigma^-$, 
do not commute with each others: therefore $H_{int}$ does not describe a 
pure-dephasing interaction.
 \noindent The proper ECS are still the Glauber coherent states, while the operators $H(\alpha)$, acting on $\Gamma$, read
\begin{equation}
    H(\alpha) = \bra{\alpha}H\ket{\alpha}=\omega |\alpha|^2 \mathbb{1} + g(\alpha \sigma^+ + \alpha^* \sigma^-) \,,
\end{equation}
and we can write
\begin{equation}
    \frac{\partial H(\alpha)}{\partial \alpha^*} = g\sigma^- + \omega\alpha \mathbb{1} \,,\qquad  \frac{\partial H(\alpha)}{\partial \alpha} = g\sigma^+ + \omega\alpha^* \mathbb{1}\,;
\end{equation}
therefore, the objects $\ket{\smash{\dot{\alpha}}}=D(\dot{\alpha})$ are obtained as
\begin{equation}
\begin{aligned}[b]
    & D(\dot{\alpha}) = \exp \Big[ -i(g\sigma^- + \omega\alpha \mathbb{1})\otimes a^\dagger -i(g\sigma^+ + \omega\alpha^* \mathbb{1})\otimes a \Big] \\
    & = \exp \Big[ -ig(\sigma^+\otimes a + \sigma^-\otimes a^\dagger ) -i\omega \mathbb{1} \otimes (\alpha^* a + \alpha a^\dagger) \Big]~.
\end{aligned}
\end{equation}
To get the operators $F_\pm,F_\pm^\dagger$, we need to 
explicitly write the action of $D(\dot{\alpha})$ on $\ket{0}$.
On the other hand, the presence of the non-commuting operators $\sigma^+$ and 
$\sigma^-$ (in contrast with the previous case, where we only had to 
deal with $\sigma_z$), makes it more difficult to factorize 
$D(\dot{\alpha})$ as a product of exponentials. Therefore, at variance 
with the pure-dephasing case, we here consider the classical limit of 
the environment right from the beginning, just to factorize 
$D(\dot{\alpha})$. In fact, it can be shown\cite{FCVpra16,Foti19} that
\begin{equation}
    D(\dot{\alpha})\ket{0}\to e^{-ig\sigma^-\otimes a^\dagger } 
e^{-i\omega\alpha\mathbb{1}\otimes a^\dagger} \ket{0}~,
    \label{eq:jc_disentangled}
\end{equation}
and hence Eq.~\eqref{eq:result_integral} reads
\begin{equation}
    \dot{\rho}_\Gamma = -i\big[H_{\rm eff}, \rho_\Gamma \big] +  \int d\mu\,{\rm T} \,\Big( \sigma^+ R \sigma^- - \frac{1}{2}\big\{\sigma^-\sigma^+ , R \big\} \Big),
\end{equation}
where
\begin{equation}
\begin{aligned}
    & {\rm T}= 2g^2 \sum_k \gamma_k |v_k|^2\, , \\
    & H_{\rm eff} = -g\sum_k \int d\mu \,\gamma_k \Big[ \Im(v_ku^*_k)\sigma_x +\Re(v_ku^*_k)\sigma_y  \Big]\, ,
\end{aligned}
\end{equation}
and $v_k,u_k$ are functions on the manifold. Getting
their expression would be the same as solving the 
exact quantum dynamics of the system and its enviroment, altogether, 
which is not an achievable task.
However, reminding Eq.~\eqref{e.chi2_classical},
when the environment becomes macroscopic and behaves classicaly
we get
\begin{equation}
    \dot{\rho}_\Gamma \approx -i[\tilde{H}_{\rm eff},\rho_\Gamma] + \widetilde{\rm T}\Big( \sigma^+ \rho_\Gamma \sigma^- - \frac{1}{2}\big\{\sigma^-\sigma^+ , \rho_\Gamma \big\} \Big)
\label{e.rho_JC_classical}
\end{equation}
for some $\widetilde{H}_{\rm eff}$ and $\widetilde{\rm T}$ that
depend neither on $\alpha$ nor on $t$, which makes the above equation 
\eqref{e.rho_JC_classical} the well known master equation for the 
Jaynes-Cummings model.

\section{Conclusions}
\label{s.V}
In this work we have 
{\it i)} considered the reduced density operator 
for an OQS, 
{\it ii)} written its parametric representation with environmental 
coherent states, and 
{\it iii)} shown that its time derivative satisfies an 
equation that can be cast into a GKSL form, once some assumptions are 
made.

In particular, we have shown that it is the time derivative of 
the distribution $\chi^2(\Omega(t))$, representing the probability for 
the environment to be in the ECS labelled by the parameter 
$\Omega(t)$, that generates a GKSL structure. Moreover, 
the operators $F_k, F_k^\dagger$ that we recognize as the siblings 
of the usual lindbladians, are defined locally on 
the manifold $\mathcal{M}$, that is, every $\Omega\in\mathcal{M}$ has 
associated to it a set of lindbladians $\{F_k(\Omega)\}$ entering the 
equation for $\dot{\rho}(t)$. Such operators depend on the microscopic 
details of the theory and, in principle, can be computed explicitly once
the total hamiltonian of the composite system is given. Clearly,
this is a highly non-trivial task, and can only be done in some specific 
cases. Let us finally underline that the use of ECS in 
our approach is crucial, as continuous parametric 
representations are not generally equipped with some inherent dynamics 
on the manifold, in the way ECS are. The possibility of associating a 
time evolution $\Omega(t)$ to environmental states $\ketom$ stems from the 
fact that the dynamics of coherent states is known, and it is of 
classical type, with a classical-like hamiltonian function that is 
related to the original hamiltonian operator. This provides an 
interpretation that proves itself appropriate to obtain a GKSL 
general structure, and motivates further 
investigation along these lines, as well as comparison with 
related recent works, such as 
Refs.~\cite{AlbertEtal16, DenisovEtal19, MalekakhlaghMG22}.


\end{document}